\documentclass[longbibliography,prr,twocolumn,superscriptaddress,amssymb]{revtex4-1}
\usepackage[english]{babel}

\usepackage[colorinlistoftodos, color=green!40, prependcaption]{todonotes}

\usepackage[colorlinks=true,linkcolor=magenta,citecolor=magenta,urlcolor=magenta,linktocpage=True]{hyperref}
\usepackage{amsthm}
\usepackage{mathtools}
\usepackage{physics}
\usepackage{xcolor}
\usepackage{graphicx}
\usepackage{adjustbox}
\usepackage{placeins}
\usepackage[T1]{fontenc}
\usepackage{lipsum}
\usepackage{csquotes}
\usepackage{pythonhighlight}
\usepackage{cleveref}
\usepackage{listings}
\usepackage{markdown}

\lstnewenvironment{numbered_python}[1][]{\lstset{style=mypython,numbers=left}}{}

\definecolor{red}{rgb}{0,0,0}

\definecolor{codegreen}{rgb}{0,0.6,0}
\definecolor{codegray}{rgb}{0.5,0.5,0.5}
\definecolor{codepurple}{rgb}{0.58,0,0.82}
\definecolor{backcolour}{rgb}{0.95,0.95,0.92}
\lstdefinestyle{mystyle}{
    backgroundcolor=\color{backcolour},   
    commentstyle=\color{red},
    keywordstyle=\color{magenta},
    numberstyle=\tiny\color{codegray},
    stringstyle=\color{codepurple},
    basicstyle=\ttfamily\footnotesize,
    breakatwhitespace=false,         
    breaklines=true,                 
    captionpos=b,                    
    keepspaces=true,                 
    numbers=none,                    
    numbersep=5pt,                  
    showspaces=false,                
    showstringspaces=false,
    showtabs=false,                  
    tabsize=2
}
\newcommand{\ex}[1]{\langle #1 \rangle}

\newcommand{\beq}{\begin{eqnarray}}
\newcommand{\eeq}{\end{eqnarray}}

\newcommand{\figref}[1]{\mbox{Fig.~\ref{#1}}}

\renewcommand{\eqref}[1]{\mbox{Eq.~(\ref{#1})}}

\newcommand{\be}{\begin{equation}}
\newcommand{\ee}{\end{equation}}
\newcommand{\bea}{\begin{eqnarray}}
\newcommand{\eea}{\end{eqnarray}}

\newcommand{\paul}[1]{{\color{red} #1}}

\begin{document}
\title{Fixing detailed balance in ancilla-based dissipative state engineering}

\author{Neill Lambert}
\email{nwlambert@gmail.com}
\affiliation{Theoretical Physics Laboratory, Cluster for Pioneering Research, RIKEN, Wakoshi, Saitama 351-0198, Japan}
\affiliation{Quantum Computing Center, RIKEN, Wakoshi, Saitama, 351-0198, Japan}
\author{Mauro Cirio}
\email{cirio.mauro@gmail.com}
\affiliation{Graduate School of China Academy of Engineering Physics, Haidian District, Beijing, 100193, China}
\author{Jhen-Dong Lin}
\affiliation{Department of Physics, National Cheng Kung University, 701 Tainan, Taiwan}
\affiliation{Center for Quantum Frontiers of Research and Technology, NCKU, 701 Tainan, Taiwan}
\author{Paul Menczel}
\affiliation{Theoretical Physics Laboratory, Cluster for Pioneering Research, RIKEN, Wakoshi, Saitama 351-0198, Japan}
\author{Pengfei Liang}
\email{pfliang@csrc.ac.cn}
\affiliation{Graduate School of China Academy of Engineering Physics, Haidian District, Beijing, 100193, China}
\author{Franco Nori}
\affiliation{Theoretical Physics Laboratory, Cluster for Pioneering Research, RIKEN, Wakoshi, Saitama 351-0198, Japan}
\affiliation{Quantum Computing Center, RIKEN, Wakoshi, Saitama, 351-0198, Japan}
\affiliation{Physics Department, University of Michigan, Ann Arbor, MI 48109-1040, USA}

\date{\today} % Leave empty to omit a date

\begin{abstract}
Dissipative state engineering is a general term for a protocol which prepares the ground state of a complex many-body Hamiltonian using engineered dissipation or engineered environments. Recently, it was shown that a version of this protocol, where the engineered environment consists of one or more dissipative qubit ancillas tuned to be resonant with the low-energy transitions of a many-body system, resulted in the combined system evolving to reasonable approximation to the ground state. This potentially broadens the applicability of the method beyond non-frustrated systems, to which it was previously restricted.  Here we argue that this approach has an intrinsic limitation because the ancillas, seen as an effective bath by the system in the weak-coupling limit, do not give the detailed balance expected for a true zero-temperature environment. Our argument is based on the study of a similar approach employing linear coupling to bosonic ancillas.  We explore overcoming this limitation using a recently developed \paul{open quantum systems technique} called pseudomodes. With a simple example model of a 1D quantum Ising chain, we show that detailed balance can be fixed, and a more accurate estimation of the ground state obtained, at the cost of two additional unphysical dissipative modes and the extrapolation error of implementing those modes in physical systems.
\end{abstract}


\maketitle

\section{Introduction}

Quantum computers are expected to be able to efficiently simulate \cite{RevModPhys.86.153} the dynamics of complex many-body quantum Hamiltonians. However, finding the ground state of such a Hamiltonian is not so easy \cite{cubitt}, and it is believed that even a fully fault-tolerant quantum computer will not be able to find the ground state of every arbitrary quantum, or even classical, Hamiltonian in less than exponential time. Nevertheless, there is the potential that a quantum algorithm can still perform this task more efficiently than any known classical algorithms. Such an algorithm would have important applications in topics ranging from quantum chemistry and quantum materials to classical optimization problems. 

One approach for finding such ground states is that of dissipative state engineering \cite{natphys, Lanyon2011, lloyd}.  In this scenario, the many-body Hamiltonian is encoded digitally on a universal quantum computer, or engineered in some physical system in an analog form, and an artificial zero-temperature environment is designed to simulate the cooling of this system to the ground state (similar approaches have been developed that, when combined with intrinsic symmetries, allow dissipation to prepare complex entangled states \cite{PhysRevLett.113.040501}).  Initial conceptualizations of this method, in applications to finding ground states, argued it could be useful also in non-error corrected quantum computers, as the dissipative process would be self-correcting. However, the initial proposal \cite{natphys} concentrated only on frustration-free systems.

Recently, {\color{red} several new methods \paul{aiming to overcome this limitation} appeared in the literature, including \paul{an approach based on weak measurement} \cite{cubitt} and \paul{one} using filter functions and Lindbladian dissipation \cite{linlinground}. A more heuristic approach using resonant ancillas as dissipative baths} \paul{has been proposed in} \cite{Cormick2013, sciadv, 9259940}, and was recently implemented experimentally in \cite{googlepublished}. In those works, one or more dissipative ancillas are coupled to one or more parts of the many-body system and the total new composite system is evolved until thermalization occurs. Ideally, as the ancillas remove energy from the system, the long time behavior of this composite system should push the original many-body system close to its ground state, even in the presence of frustration.

In \cite{sciadv}, the authors found that a qubit ancilla was efficient at cooling several many-body example systems close to their ground state, in particular the quantum Ising model and the Heisenberg model.  Optimality was found by choosing the energy of the ancilla close to the lowest energy transition of the system $E_{01}=E_1 - E_0$, where $E_0$ is the ground-state (manifold) energy and $E_1$ the first excited state. In addition, as one might intuitively expect, larger coupling and dissipation made the cooling process faster. However, using too large values inevitably reduced the efficiency of the approach due to hybridization and Zeno effects, respectively. 

In optimal cases, for the examples studied therein, a fidelity of around $90$\% was observed.  What is the intrinsic limitation at play? It is useful to explicitly look at the bosonic-ancilla variant of this approach to understand what this limitation might be, and to overcome it. The coherent dynamics of the composite system and bosonic ancilla are given by a total Hamiltonian
\beq
H = H_{\mathrm{s}} + H_{\mathrm{p}} + H_{\mathrm{int}} ,\label{1}
\eeq
where $H_{\mathrm{s}}$ is the many-body system Hamiltonian,  $H_{\mathrm{p}} = \omega_1 a_1^{\dagger} a_1$ is the ancilla Hamiltonian which we write as a single bosonic mode, and $H_{\mathrm{int}} = \lambda_1 Q \left(a_1+a_1^{\dagger}\right) $ is the interaction involving an operator $Q$ on the space of the system, and $\lambda_1$ is a coupling strength. Note that our use of linear coupling to the bosonic ancilla is not completely equivalent to the qubit-ancilla approach used in \cite{sciadv}, even in the low-excitation limit, because of the more generalized bath-coupling operator they employ therein.  

The dissipative \paul{dynamics} of the ancilla is given by a single Lindblad dissipator which removes energy from the ancilla mode at a rate $\gamma_1$ using just local operators acting on the ancilla itself,
\beq
L[\rho] = \gamma_1 \left[a \rho a^{\dagger} - \frac{1}{2}\left(a^{\dagger}a \rho + \rho a^{\dagger}a\right) \right], \label{2}
\eeq
such that the equation of motion of the composite system is a standard ``operator-local'' Lindblad master equation $\dot{\rho}=-i[H,\rho] + L[\rho]$. 

Returning to the question of what limits this kind of dissipative-ancilla-based cooling, with this bosonic formulation it is fairly straightforward to draw an analogy between the dissipative ancilla and weak coupling to a structured environment. The standard theory of open quantum systems tells us that normally the steady state is determined by the detailed balance condition \cite{Alicki1976,Kossakowski1977,Frigerio1990,RevModPhys.82.1155}, i.e., the balance between the power spectrum of the environment at positive and negative frequencies, $S(\omega) = \exp \left[\beta \omega\right]S(-\omega)$, where $\beta = 1/k_{\mathrm{B}}T$ is the inverse temperature (here we use $k_{\mathrm{B}} = \hbar = 1$ throughout).

The power spectrum of the ancilla presented above can be found as 
$S(\omega) = \int_{-\infty}^{\infty}e^{i\omega t}C(t)\; dt$, 
% $S(\omega) = \int_{-\infty}^{\infty}e^{i\omega t}C(t)$, 
where $C(t)=\ex{\mathbb{X}(t)\mathbb{X}(0)}$ is the correlation function of the free-bath coupling operator $\mathbb{X} = \lambda_1(a+a^{\dag})$ around the free-bath steady-state.  For the single damped \paul{bosonic} mode defined above, \paul{the result} is,  at zero temperature, simply a  Lorentzian
$S(\omega)=  2 \lambda^2_1 \gamma_1/[(\omega-\omega_1)^2+\gamma_1^2]$.

To achieve a steady state close to the ground state \paul{it is required that}
\beq 
S(-\omega)=0, \label{sw0}
\eeq  but this \paul{requirement} is clearly not \paul{satisfied for this Lorentzian} power spectrum of the dissipative ancilla {\color{red} whose \paul{dissipative dynamics} is given by \eqref{2}. Instead,} the system sees an effective frequency-dependent temperature %$T_{\mathrm{eff}}(\omega)=\omega/\mathrm{log}\left[\{(\omega_1+\omega)^2+\gamma_1^2\}/\{(\omega_1-\omega)^2+\gamma_1^2\}\right]$ 
\beq
T_{\mathrm{eff}}(\omega)=\omega/\mathrm{log}\left[S(\omega)/S(-\omega)\right]\eeq (see \figref{fig6} for an example). Note however that with such a frequency-dependent \paul{effective} temperature, a Gibb's-like distribution is not guaranteed \paul{in the steady state}. 
Nevertheless, this effective temperature $T_{\mathrm{eff}}$ serves as a useful goal function to minimize when designing a bath which will take a system close to its ground state.

\begin{figure}[t!]
\includegraphics[width = \columnwidth]{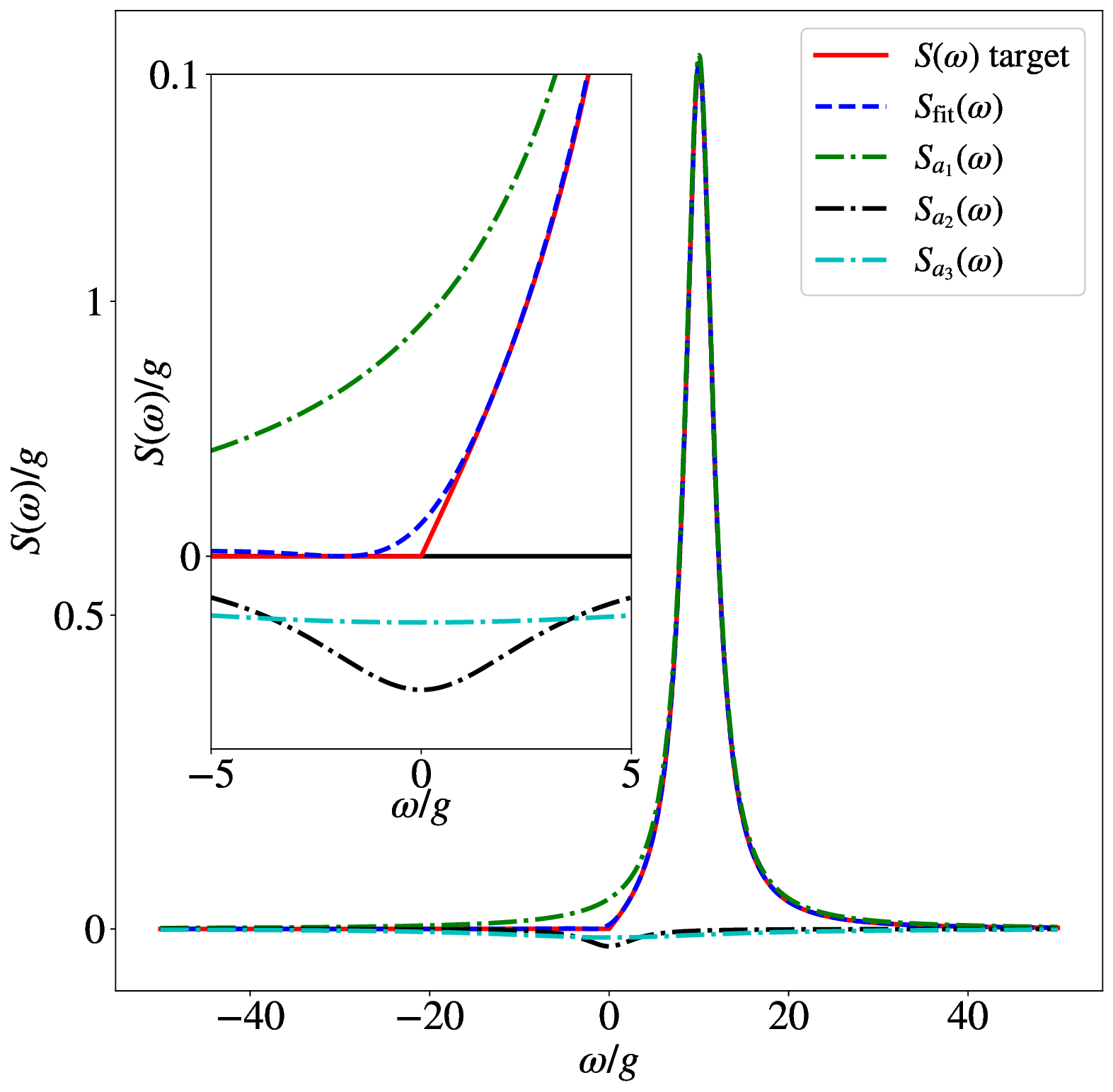} 
\caption{The power spectrum of a zero temperature dissipative discrete quantum system does not directly correspond to that of a true zero temperature environment. For the purposes of dissipative state engineering, this limits its ability to bring a complex many-body system to its ground state. 
 To compensate, here we take advantage of the ability of pseudomodes to contribute negative terms to an effective total power spectrum, and suppress the negative frequency domain which causes unwanted heating. 
 Here we show the target power spectrum extracted from $S(\omega) = 2 J(\omega)[n_{\mathrm{th}}(\omega) +1]$, the three effective power spectra of the ancillas, and sum of the three $S_{\mathrm{fit}}(\omega)=S_{a_1}(\omega)+S_{a_2}(\omega)+S_{a_3}(\omega)$, giving the fit to $S(\omega)$. The inset shows a zoom around zero frequency, and parameters are those used in \figref{2}. \label{fig1}}
\end{figure}

\section{Fixing detailed balance} 

As discussed in  \cite{lambert,PhysRevA.55.2290,Tamascelli,PhysRevResearch.2.043058, Cirio2022,Dorda,PleasanceArXiv210805755Quant-Ph2021,TrivediPhysRevLett2021}, fixing this broken detailed balance is difficult using additional dissipative ancillas alone, as they essentially add more positive-valued Lorentzians to a given power spectrum.  In fact, fixing detailed balance requires the addition of negative terms to the total power spectrum, not more positive contributions. Fortunately, this is exactly what is provided by the powerful pseudomode approach. This method allows us to consider
ancillas which generate negative valued correlation functions,  thereby adding Lorentzians  with negative amplitude to the effective power spectrum. This allows us to negate unwanted positive contributions as needed and to generate \paul{a power spectrum that is} as close as possible to be zero-valued at negative frequencies, as per \eqref{sw0} and as demonstrated in \figref{1}.

The details of this method have been discussed elsewhere  \cite{lambert, Luosi, Cirio2022, Menczel2023}, \paul{and we} will just summarize the main results needed for this work \paul{in Appendix \ref{pm_appendix}}.  We assume that the pseudomode method is attempting to model a true continuum environment at zero temperature with a spectral density given by 
\beq 
J(\omega) = \lambda^2 \gamma \omega/\left[(\omega^2-\omega_0^2)^2 + \gamma^2 \omega^2\right].\eeq
As shown in \cite{lambert}, this generates a bath correlation function composed of a resonant term and an infinite sum of Matsubara frequencies which can either be fit by two exponentials and represented by two additional ancillas, or encoded  by an imaginary-valued stochastic field \cite{Luosi}.  For simplicity, here we focus on the former, but the latter approach can also be applied, reducing the number of ancilla modes at the cost of averaging  over the classical noise.

Given the decomposition discussed in  \cite{lambert}, we acquire a pseudomode model of the original system coupled to three ancillas
\beq
H_{\mathrm{tot}} = H_{\mathrm{s}} + H_{\mathrm{pm}} + H_{\mathrm{s-pm}}
\eeq
where $H_{\mathrm{s}}$ is the many-body system Hamiltonian,  $H_{\mathrm{pm}} = \sum_{j=1}^3\omega_j a_j^{\dagger} a_j$, and \begin{equation}
\label{eq:Hpms}
H_{\mathrm{s-pm}} =  Q\left[\lambda_1\left(a_1+a_1^{\dagger}\right) +\bar{\lambda}\sum_{j=2}^3\lambda_j\left(a_j+a_j^{\dagger}\right) \right]   \;  
\end{equation}
is the interaction between the system and pseudomodes.
\paul{Here,} $Q$ is an operator on the space of the system, and $\lambda_j$ is the coupling strength between the system and pseudomode $j$. \paul{The parameter $\bar\lambda$ has the value $i$}, causing the Hamiltonian to be non-Hermitian. Later we will use this parameter to analytically continue \paul{results obtained using} physical couplings \paul{($\bar\lambda \in \mathbb R$)} to the complex plane \cite{antimodes}.

As before, the dissipative \paul{dynamics} of the ancilla \paul{is based on only local jump} operators,
\beq
L[\rho] = \sum_{j=1}^{3}\gamma_j \left[a_j \rho a_j^{\dagger} - \frac{1}{2}\left(a_j^{\dagger}a_j \rho + \rho a_j^{\dagger}a_j\right) \right]. \label{local}
\eeq
Now the equation of motion of the composite system  $\rho_{\mathrm{tot}}$ is a time-local and operator-local equation $\dot{\rho}_{\mathrm{tot}}=-i[H_{\mathrm{tot}},\rho_{\mathrm{tot}}] + L[\rho_{\mathrm{tot}}]$, but it is technically not a Lindblad master equation because $\bar{\lambda}$ is evaluated at an imaginary value.   
In addition, it is sometimes termed a  pseudo-Lindblad equation to account for the fact that complex conjugation of the complex Hamiltonian is not performed.

Importantly, the bath spectral density is chosen so that the first ancilla, $a_1$, has properties akin to those of the qubit ancilla defined in \cite{sciadv}. As discussed earlier, on its own, it would induce an incorrect detailed balance and the system would see a non-zero effective temperature. By introducing \paul{the additional ancillas} $a_2$ and $a_3$, detailed balance is (partially) restored and, together, these modes can induce  the system to relax  to a state closer to  its true ground state.  

However, several issues do remain that prevent this method from obtaining perfect ground state fidelity. First, the use of a fit to obtain the properties of $a_2$ and $a_3$ necessarily induces a residual error in the effective power spectrum.  For our purposes, this is detrimental,  as
it amounts to \paul{unwanted small but non-zero} values of $S(\omega)$ for $\omega < 0$. {\color{red} This can be further mitigated by the addition of more pseudomodes. The convergence of this error, in the number of modes, depends on the fitting approach employed~\cite{10.1063/5.0209348}, and the optimality of different methods will be explored in future work. Second, these modes do not act like a perfect weak-coupling bath, and hybridization can still induce a finite error (we will explore an approach to mitigate this later). Finally,} while the operator-local nature of \eqref{local} [compared to a global master equation, see \eqref{bms}] is an important feature for its practical implementation \cite{lloyd, 9259940, googlepublished} in quantum hardware (e.g., by being  emulated with additional ancillas which are periodically reset \cite{lindbladsim1, lindbladsim2}), {\color{red} a physical implementation of the system's interaction with the modes $a_2$ and $a_3$ cannot be done directly, \paul{since the interaction is unphysical by definition}}. Fortunately, this \paul{issue} can be circumvented by analytical continuation of results obtained from purely physical models \cite{Iblisdir,self2022estimating,antimodes}.

\begin{figure}[t!]
\includegraphics[width = \columnwidth]{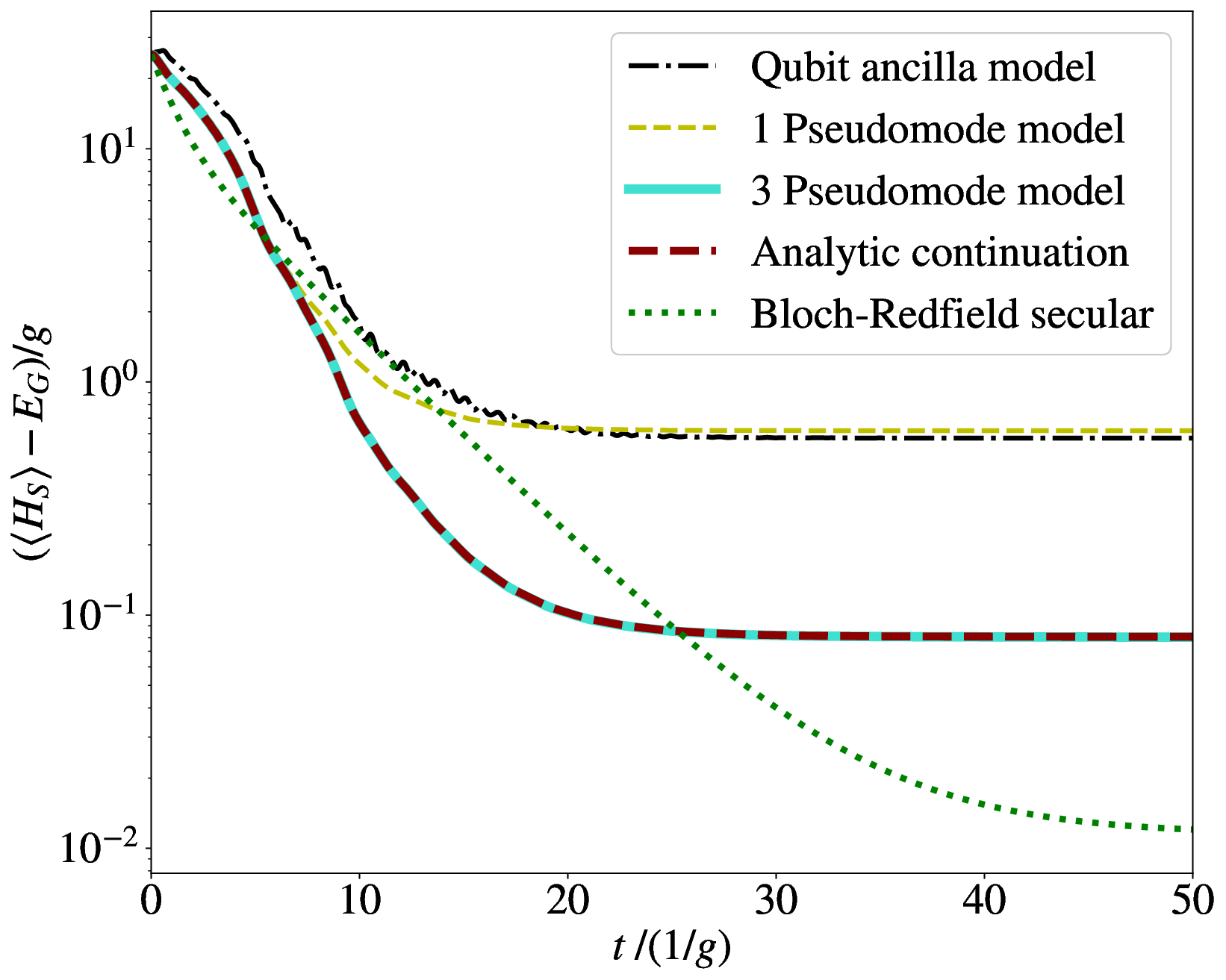} 
\caption{For the Ising model in \eqref{ising} we show results for $N=5$, $J=5g$, and total effective bath spectral density parameters  $\omega_0 = 1.2 E_{01}$, $\gamma = 3.8g$, and $\lambda = 1.15g \sqrt{\Omega}$ with $\Omega = \sqrt{\omega_0^2 - (\gamma/2)^2}$.  We see that the qubit ancilla model (with more generalized couplings, see \cite{sciadv}) and the single bosonic ancilla model are both similarly limited in their ability to minimize the energy difference $(\langle H_S \rangle - E_G)/g$ where $E_G$ is the ground-state energy  of the of the Ising model.  The three-mode pseudomode model (solid turquoise curve) provides a better result, and, for intermediate times, faster than the equivalent Bloch-Redfield solution, \eqref{bms}, using $S_{\mathrm{fit}}(\omega)$ (dashed green curve).  The dark red, dashed curve shows the result of a simulation using extrapolation from a purely \paul{physical} pseudomode model. This is based on fitting to solutions obtained using nine different real values for the coupling $\bar{\lambda}$, and extrapolation with an order six polynomial.\label{fig2}}
\end{figure}

\begin{figure}[t!]
\includegraphics[width = \columnwidth]{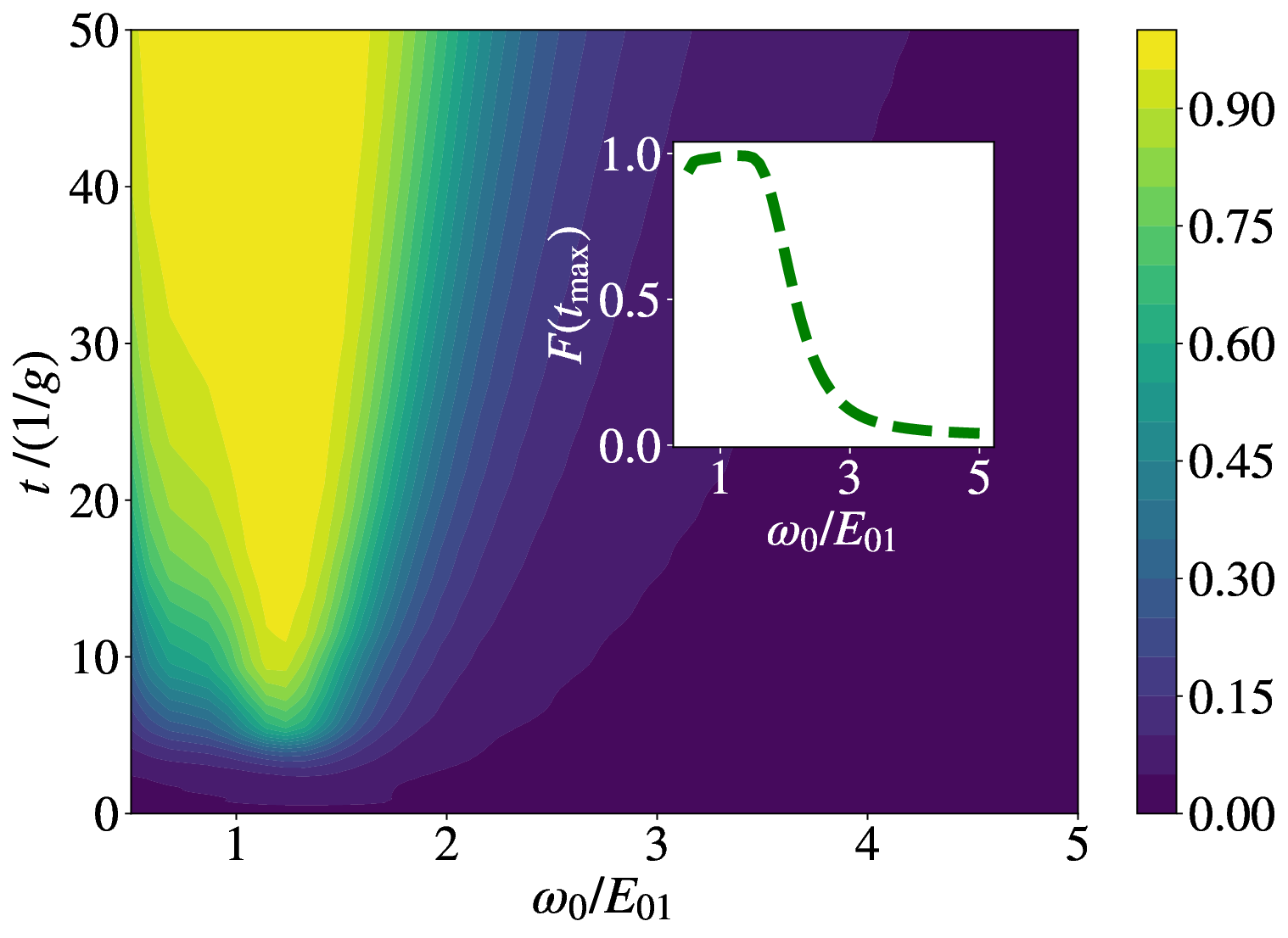} 
\caption{Here we show, for the full pseudomode model, the fidelity with respect to the ground state of the Ising model as a function of time and bath resonance frequency $\omega_0$, using the same parameters as in \figref{fig2}.  The inset shows the fidelity at the maximum time evolved for in the contour plot, $t_{\mathrm{max}}=50$ $(1/g)$.  In contrast to a single mode (which reaches around $90\%$ fidelity), the full model obtains high fidelity for a range of $\omega_0$ around $E_{01}$, reaching a maximum slightly off-resonance due to the corresponding reduced hybridization (see \figref{fig6} for more details).  For large detunings, it takes longer to reach the same target fidelity due to the lower effective dissipation rates induced by the ancillas. \label{fig3}}
\end{figure}

To demonstrate the utility of this approach, we now focus on the main example provided in \cite{sciadv}:  an Ising spin chain where the effective bath is coupled to a single spin at the end of the chain.  The Ising model, which constitutes the many-body system, is given by,
\beq
H_{\mathrm{s}} = g\sum_{j=1}^N \sigma_z^{(j)} - J\sum_{j=1}^{N-1} \sigma_x^{(j)}\sigma_x^{(j+1)}.\label{ising}
\eeq
In traditional terminology, $g$ is the transverse field and $J$ is the coupling strength.  A paramagnetic phase $g>J$ is separated from a ferromagnetic phase $g<J$ by a critical point at $g=J$. We couple this many-body system via one edge spin to the pseudomode environment defined above, using the operator $Q = \sigma_x^{(N)} + 1.1 \sigma_y^{(N)} + 0.9 \sigma_z^{(N)}$ (adapted from \cite{sciadv}, where it was chosen to break symmetries in the system),
and compare it to  both the single-qubit ancilla results found in \cite{sciadv} and the single bosonic ancilla, i.e.,  without the additional $a_2$ and $a_3$ modes. We focus on the $J>g$ phase first, and use parameters close to those outlined in \cite{sciadv}, to find the dynamics and steady-state properties of the average energy and ground-state fidelity. As in that work, when the ground state is degenerate, we consider a projection onto that subspace  to define the fidelity. {\color{red}\paul{Note that the symbol $E_{01}$ in figures \ref{fig2} and \ref{fig3} and  refers} to the gap between that subspace and the first excited state relative to it.}

 First, in \figref{fig1}, we plot the target power spectrum $S(\omega)$ and the effective power spectra generated by the three ancillas used in the pseudomode model which together form $S_{\mathrm{fit}}(\omega)$.  The two unphysical modes $a_2$ and $a_3$ contribute negative-valued power spectra and thus correct the detailed balance of the single resonant mode $a_1$.

 Next, in  \figref{fig2}, we show the dynamics \cite{Qutip1,Qutip2} of the single-qubit ancilla model, single-bosonic ancilla model, and the full pseudomode model. The \paul{steady state of the full pseudomode model approximates the true ground state energy more accurately}, with a fidelity of around $99\%$. This is shown more clearly in \figref{fig3}, where we find optimal parameters very close to those used for the single-qubit ancilla, but at a bath resonance frequency slightly higher than the system's low-energy gap $E_{01}$.
 
 As mentioned, residual deviations from the \paul{true} steady state can arise from hybridization and fitting errors. This can be verified by comparing to a solution obtained by a Bloch-Redfield master equation~\cite{Petruccione, Lidar} %[\eqref{bms} using $S_{\mathrm{fit}}(\omega)$], 
 as shown in Fig.~\ref{fig2} (see \figref{fig4} in the appendix for another example). Typically, the hybridization error can be seen in the difference between the Bloch-Redfield and pseudomode steady-state results, while the fitting error is demonstrated by the difference in the steady-state Bloch-Redfield and true ground-state energy. In \figref{fig2}, we see that the hybridization error dominates [the Bloch-Redfield results saturate around $t=100$ $(1/g)$, at $(\langle H_S \rangle - E_G)\approx 10^{-2}g$]. The hybridization can be minimized by reducing $\lambda$ at the cost of a longer cooling time, while the fitting error can be reduced with additional ancillas.

The parameter regime used in \figref{fig2} and \figref{fig3} involves  a two-fold degeneracy of the
ground state subspace of the Ising model, and a very large energy gap to the first excited state. A more challenging regime is  for $J=1.4g$, which is closer to the critical point of this model, and where the energy gap is much smaller.

If we still classify the two-lowest, now non-degenerate, energy states as the target subspace, we see results comparable to \cite{sciadv}; but if we really want to distinguish the two lowest-lying states, it is more challenging.  The small gap limits the magnitude of bath couplings we can use \paul{while avoiding} hybridization. This means that we need a much longer cooling time to achieve a high fidelity result.  To avoid this, we can take advantage of the controllable nature of the artificial environment. In \figref{fig4},  an initially large coupling strength is used to
drive the system close the ground state. Then, a secondary quench of the coupling strength to a much weaker value is made, to eliminate residual excitations due to hybridization with the bath. This suggests the combination of ancilla-based dissipative state-engineering with other state-engineering approaches, like adiabatic methods \cite{Aharonov2008}, or variational techniques \cite{troyer,Stanisic2022}, might also be viable.

\section{Analytical continuation.} 

To implement pseudomodes in real experiments, either via analog systems or digitally with ancillas and measurements \cite{googlepublished,preskil}, the pseudomodes themselves must have real physical parameters.  They demonstrably do not, so at first glance what gives them their ability to efficiently represent a physical bath also makes them fundamentally ``unphysical'' and unfit for this purpose.

However, in a recent work \cite{antimodes} we demonstrated how the results of the general unphysical pseudomode model could be obtained from a physical model via analytical continuation (see also \cite{Iblisdir,self2022estimating}). {\color{red} In essence, this can be done by \paul{repeating} the quantum simulation of our model with \paul{several ($N_s$) different choices of only physical couplings}. One then extracts the observable of interest, \paul{such as the ground state energy,} from these $N_s$ simulations. \paul{From this data, the observable as a function of the real-valued coupling parameters can be fitted} with an appropriate fitting function, and \paul{then extrapolated} into the complex-coupling domain.  

How far one needs to extrapolate into the complex plane determines the range of \paul{real} values of couplings that one needs to implement. In addition, the complexity of the functional dependence of the desired observable on the coupling parameter determines the \paul{required} order of the fitting polynomial $M$, which in turn determines how many ($N_s>M$) experiments one must perform. }

%^real physical systems for $N_s$ real-valued choices of the coupling strengths $\bar{\lambda}$, fitting the resulting observables, like the ground-state energy, and extrapolating to the complex-valued coupling required in Eq.~(\ref{eq:Hpms}). 

In \figref{fig2}, we demonstrate an extrapolation of the ground state energy as the observable of choice. In this example, we solve the three-mode pseudomode model with purely real coupling strengths {\color{red} between the system and} modes 2 and 3 for a set of $N_s=9$ equally spaced values for the parameter 
 $\bar{\lambda}\in [0,1]$ in Eq.~(\ref{eq:Hpms}), and then extrapolate to the desired unphysical pseudomode model at $\bar{\lambda}\rightarrow i$ using a sixth order polynomial.  We note that this procedure is only possible because of the unusual nature of the pseudo-Lindblad equation which does \emph{not} enforce Hermitian conjugation on its complex parameters.

In \figref{fig2}, we see an almost perfect result for the extra\-polation. However, understanding the general errors that may arise from this approach is not trivial. In \cite{antimodes}, we found that the total error of such an extrapolation consists of a bias error (how precisely the polynomial of order $M$ captures the true functional form of the data), and a stability error (due to the finite knowledge we can obtain about given observable in an experiment, i.e., the measurement error). Importantly, the bias error decreases with $M$, while the  number of measurements needed to suppress the stability error grows exponentially in $M$, leading to an interesting trade off between these two quantities.

As a consequence, the general viability of this protocol relies on how the polynomial order $M$ (needed to obtain a result with a particular accuracy) depends on the size $N$ of our system.  Largely, this  is determined by the complexity of the extrapolated observable as a function of $\bar{\lambda}$, and thus may be problem dependent. For the example used in this work, we demonstrate this dependence numerically in \figref{fig5}, and find non-monotonic behavior in the fitting error of the ground state energy as a function of $N$ for larger values of $M$.  Interestingly, the relative error of the ground state energy actually decreases with $N$, suggesting that some extensive quantities are easy to obtain with this fitting procedure.

{\color{red} In addition, \cite{antimodes} provides a loose upper bound on the error of the fitting procedure {\color{red}in the limit of large $N_s$,} which scales linearly as a function of the system evolution time {\color{red}(multiplied by the frequency $\sum_j \lambda_j^2/\gamma_j$, see Appendix E1 in \cite{antimodes})}. This result implies that in the worst case, if the thermalization time is too long, we need a more complex fitting function (i.e., a higher order polynomial $M$), which then implies that we needed higher precision in the measurements to obtain the data. In these cases, our method of achieving faster thermalization using time-dependent couplings may be employed to speed up the overall thermalization time, see \figref{fig4}.}

    In addition to this extrapolation procedure, there is an additional cost in digital simulations \cite{https://doi.org/10.48550/arxiv.2212.08546} that arises from using bosonic instead of qubit ancillas due to the increased Hilbert space size. In the examples we show here, the resonant ($a_1$) mode is truncated at three Fock states, while the unphysical modes ($a_2$ and $a_3$) are truncated at just two Fock states, indicating the demands in the weak coupling regime are minimal. {\color{red} The two modes truncated with two Fock states can be implemented with additional single-qubit ancillas, while the mode truncated with three Fock states can be implemented using the symmetric subspace of two qubit ancillas.

    For example, {\color{red} we can consider} the correspondence $\ket{0} = \overline{\ket{00}}$, {\color{red}$\ket{1} =\left[\overline{\ket{01}}+\overline{\ket{10}}\right]/\sqrt{2}$}, and $\ket{2} = \overline{\ket{11}}$, where bare states correspond to the original Fock states, and {\color{red}overbar} states  {\color{red}belong to the qubits'} local basis.
    The Holstein-Primakoff transformation provides a simple recipe for equating operators in these two spaces. Evaluating it explicitly for this case {\color{red} allows us to approximate} the bosonic annihilation operator as, 
    \beq
    a_1 &=&  \left[\frac{1}{\sqrt{2}}\ket{0}\bra{0} +  \ket{1}\bra{1} \right]\left(\sigma_-^{(1)} + \sigma_-^{(2)}\right) \\
    &=& \frac{1}{\sqrt{2}}\overline{\ket{00}}\left(\overline{\bra{01}}+\overline{\bra{10}}\right) +  \left(\overline{\ket{01}} + \overline{\ket{10}}\right)\overline{\bra{11}} \nonumber\;.
    \eeq
    The interaction term we need to implement in a quantum simulation is our system coupled to a quadrature of this mode, $X=a_1+a_1^{\dagger}$.  In a digital simulation, where Hamiltonian dynamics are implemented with Troterrization, this term will involve three-qubit gates, whose ease of implementation depends on the underlying hardware \cite{PhysRevLett.129.220501}.
    }
    
    Finally, the optimal parameters found in \figref{fig2} and \figref{fig3} \paul{indicate} that we need coupling strengths smaller than the system's low-energy gap, and a spectral density peak slightly off-resonant from the that same gap. To find this optimal choice of resonance frequency, following \cite{sciadv}, one can take advantage of the fact that this choice also gives the fastest cooling time (see \figref{fig3}), implying that a simple optimization procedure can be applied to find this parameter.

\section{Discussion and Conclusions.} The potential use of artificial environments in quantum information is becoming a new avenue of research, allowing for novel approaches to state engineering \cite{sciadv} and error correction \cite{PhysRevLett.131.050601}.  Here we showed that the use of a single bosonic ancilla is fundamentally limited by its broken detailed balance, but this can be corrected by the use of additional unphysical ancillas without needing knowledge of the system eigenstates. We demonstrated this with a standard quantum Ising model, and verified that it can be implemented with real physical systems and extrapolation.   

There remain open questions regarding the limits of what local coupling to an ancilla can achieve \cite{PRXQuantum.3.020319,preskil,PhysRevE.50.888,Bao2021,Bao2019,Kaneko2020,Alishahiha2023} {\color{red} (e.g., systems which obey the eigenstate-thermalization-hypothesis are hypothesized to have local operators whose off-diagonal matrix elements are exponentially suppressed \cite{Moudgalya2022,PhysRevE.50.888}, see \paul{Appendix \ref{appA}} for more details)}, and about the growth of errors in the extrapolation technique for general problems. The latter adds to the resource cost of implementing this method on actual quantum hardware. For the example studied here, we showed that the error in fitting the ground-state energy showed non-monotonic behavior in $N$, potentially limiting the required polynomial \paul{degree} $M$, and thus limiting the measurement overhead in a given experiment. 

However, in general, the feasibility of the extrapolation procedure ultimately depends on the functional form of the observable.  In addition, extracting many observables, like those defining the full quantum state~\cite{aaronson1, brandao}, is much more challenging than a single extensive observable like the ground state energy. On the other hand, if the purpose of the dissipative protocol is to generate highly entangled states as input for some other algorithm, one could consider incorporating that algorithm into the ensemble of physical simulations, and only perform analytical continuation on the final outputs, when needed.

\begin{acknowledgements}
 N.L.~acknowledges  the Information Systems Division, RIKEN, for the use of their facilities, and support from the RIKEN Incentive Research Program and from MEXT KAKENHI Grant Numbers JP24H00816, JP24H00820. M.C. acknowledges support from NSFC (Grant No.~11935012). F.N. is supported in part by: Nippon Telegraph and Telephone Corporation (NTT) Research, the Japan Science and Technology Agency (JST) [via the Quantum Leap Flagship Program (Q-LEAP), and the Moonshot R\&D Grant Number JPMJMS2061], the Asian Office of Aerospace Research and Development (AOARD) (via Grant No. FA2386-20-1-4069),  and the Office of Naval Research (ONR) Global (via Grant No. N62909-23-1-2074). PM performed this work as an International Research Fellow of the Japan Society for the Promotion of Science (JSPS).
 \end{acknowledgements}
%\bibstyle{apsrev4-1}
%\bibliographystyle{apsrev}

\newpage
\appendix
\section{Born-Markov-Secular master equation} \label{appA}

In the limit of very weak coupling, we expect the effective environment described by the pseudomodes to be equivalent to that describable by a Born-Markov-Secular (BMS) master equation. Typically, in deriving such a master equation, one must take a secular Bloch-Redfield like approach, also sometimes termed a `global master equation' approach, that relies on diagonalizing $H_{\mathrm{s}}$ and defining collapse operators that connect eigenstates with rates proportional to the bath power spectrum \cite{Petruccione, Lidar},
\beq
\dot{\rho}_s(t) &=& -i\left[H_{\mathrm{s}},\rho_s(t)\right] \nonumber \\ 
&+&\sum_{i,j>i}S(\Delta_{j,i})c_{i,j} L[d_{ij}]\rho_s(t) \nonumber \\ 
&+&\sum_{i,j>i}S(-\Delta_{j,i})c_{i,j} L[d^{\dagger}_{ij}]\rho_s(t) , \label{bms}
\eeq
where, for simplicity, we have assumed no degeneracies in the eigenstates of $H_{\mathrm{s}}$ (this can be generalized easily to factor in degenerate subspaces, but leads to a more opaque notation, so we omit it here), and neglect the Lamb-shift term (which we verified has little influence), and the zero-frequency dephasing term.  Also, $d_{ij}=\ket{\psi_i}\bra{\psi_j}$ is an annihilation operator connecting eigenstates of $H_{\mathrm{s}}$, $c_{i,j}=|\bra{\psi_i}Q\ket{\psi_j}|^2$ is the matrix element connecting states $i$ and $j$ through the system operator, $Q$, that couples to the bath, and $\Delta_{i,j} = E_j-E_i$.  Moreover $L[x]\rho = x\rho x^{\dagger}-\frac{1}{2}\left[x^{\dagger}x\rho + \rho x^{\dagger}x\right]$ is the standard Lindbladian.  

For a physical environment, one typically has $S(\omega) = 2 J(\omega)\left[n_{\mathrm{th}}(\omega)+1\right]$, where $J(\omega)$ is the bath spectral density  (assumed antisymmetric in $\omega$) and $n_{\mathrm{th}}(\omega)$ is the Bose-Einstein distribution $1/[\exp(\beta \omega)-1]$.  Together with the detailed-balance condition, these imply that the steady-state is the thermal Gibbs distribution $\rho_{\th}=\exp(-\beta H_{\text{s}})/Z$, where $Z$ is the partition function which gives normalization.

This BMS master equation on its own does not help us construct a practical protocol, as we need to diagonalize $H_{\mathrm{s}}$ to construct the collapse operators, which defeats the point of a dissipative state engineering algorithm. Hence, initial attempts were restricted to frustration-free problems, where one only needed to diagonalize small local Hamiltonians to obtain a dissipator that guarantees the global ground state as the steady state \cite{natphys}. 

Nevertheless,  understanding the limitations of cooling processes as described by this master equation can be useful to understand some of the limitations of methods like the pseudomode approach.
 For example, in correcting detailed balance the pseudomode ancilla-based approach is attempting to mimic this Bloch-Redfield global Born-Markov-Secular (BMS) master equation description of an environment, and evolve the system to its groundstate (but without the need to know the system's eigenstates that a direct implementation of \eqref{bms} would need). 
 
 In practice, pseudomodes are not doing this exactly, as in the parameter regimes we employ the BMS approximation is partially violated, and we see faster cooling rates, at intermediate times, than a BMS alone would imply (see Fig.~2 in main text). However, it is useful to consider the limits of the Bloch-Redfield master equation itself. The cooling rates of such an equation are set by the power-spectrum of the effective environment, and the matrix-elements of the local operators that couple to the environment.   
The overlap between local operators and energy eigenstates is typically expected to diminish with increasing $N$,  reducing the speed at which cooling occurs \cite{PRXQuantum.3.020319}, or causing cooling protocols to get stuck in local minima \cite{preskil}. 

For example, systems obeying the eigenstate-thermalization-hypothesis may have local operators whose off-diagonal matrix elements are exponentially suppressed \cite{Moudgalya2022,PhysRevE.50.888}, \beq c_{i,j}=|\bra{\psi_i}Q\ket{\psi_j}|^2 \propto \Omega(E)^{-1/2}, \eeq where $\Omega(E)$ is the density of states at energy $E=(E_j+E_i)/2$. This quantity can potentially scale as a function of the size of the Hilbert space $2^N$ in the middle of the system spectrum, implying, in the worst case, an exponential suppression of the timescale of the cooling process relying on coupling to local operators for some classes of Hamiltonians \cite{Bao2021,Bao2019,Kaneko2020,Alishahiha2023}. Whether this directly affects  ancilla-based cooling schemes, which is more influenced by band-edges and for which the Born and Markov approximations do not directly hold, remains an interesting avenue for extending this type of approach.
\begin{figure}[t!]
\includegraphics[width = \columnwidth]{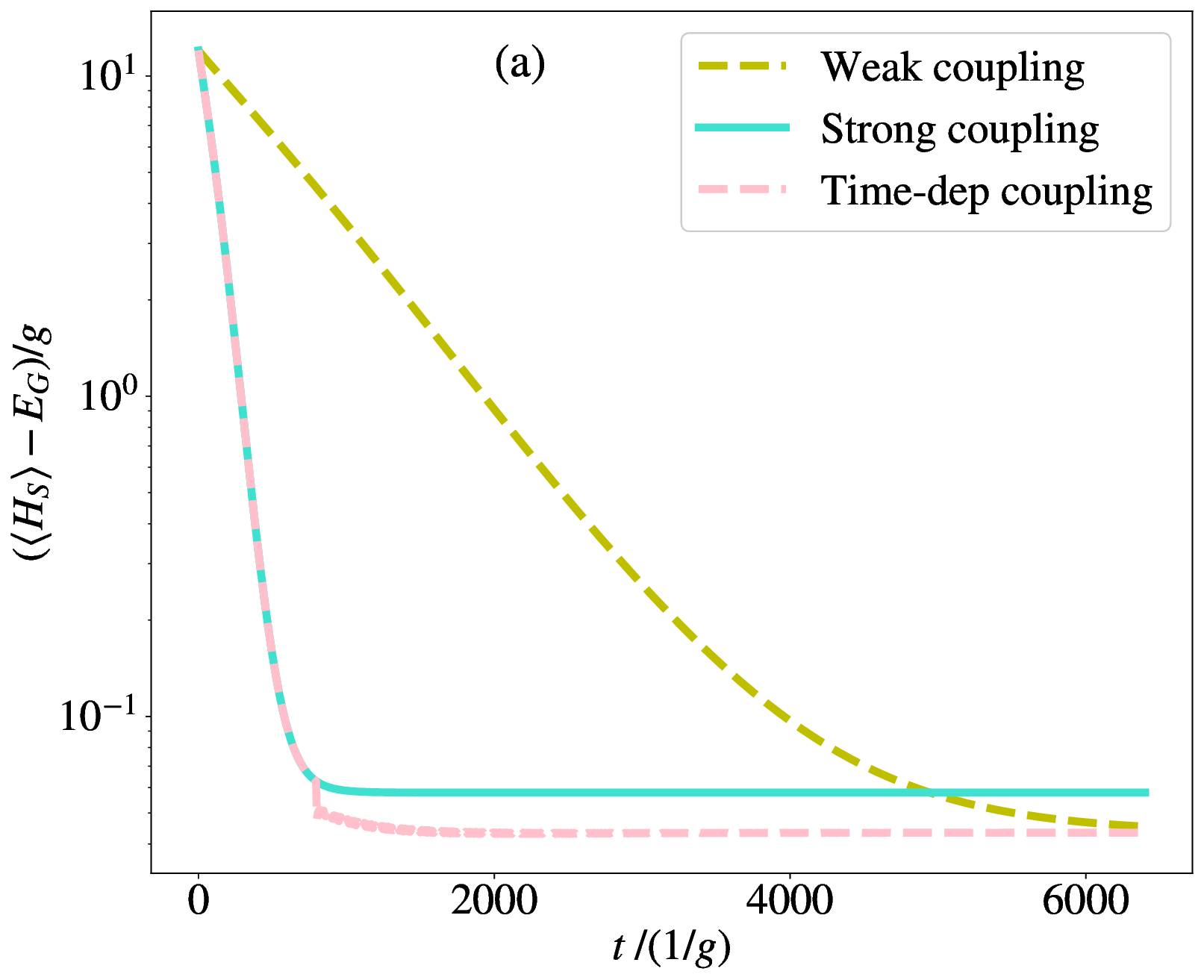} 
\includegraphics[width = \columnwidth]{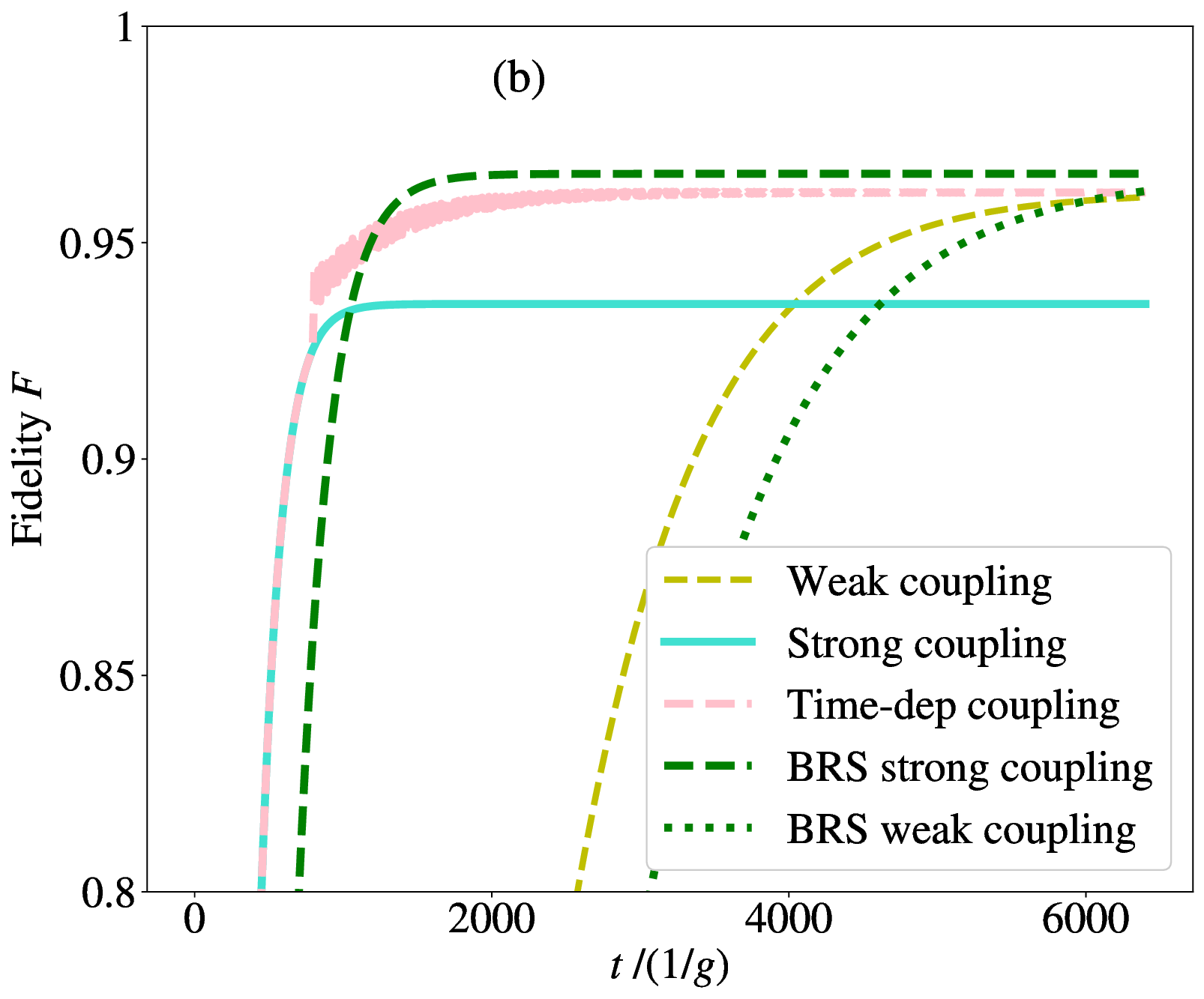} 
\caption{For the Ising model used in the main text, we present results for $J=1.4g$, and total effective bath spectral density parameters $\omega_0 = 4 E_{01}$, $\gamma = E_{01}$.  The curves labelled ``strong-coupling'' refer to a coupling $\lambda = 0.23g \sqrt{\Omega}$, where $\Omega = \sqrt{\omega^2 - (\gamma/2)^2}$, while curves labelled ``weak-coupling'' use  $\lambda = 0.11g \sqrt{\Omega}$.  The curve labelled time-dependent refers to a case where strong coupling is used up until $t=800 (1/g)$, and is then switched to weak coupling. The curves labelled BR are obtained from a Bloch-Redfield master equation simulation \eqref{bms} using the same power spectrum fit used to inform the pseudomode models. We see the steady-state of both weak and strong examples using the Bloch-Redfield approach demonstrate the same residual error as the full pseudomode model, illustrating the origin of this error is coming from the power spectrum fit, not hybridization (which is not included in the Bloch-Redfield model, compare to \figref{fig6}). This error can be further mitigated by including additional pseudomode ancillas, to achieve a better fit.\label{fig4}}
\end{figure}

\begin{figure*}[t!]
\includegraphics[width = \columnwidth]{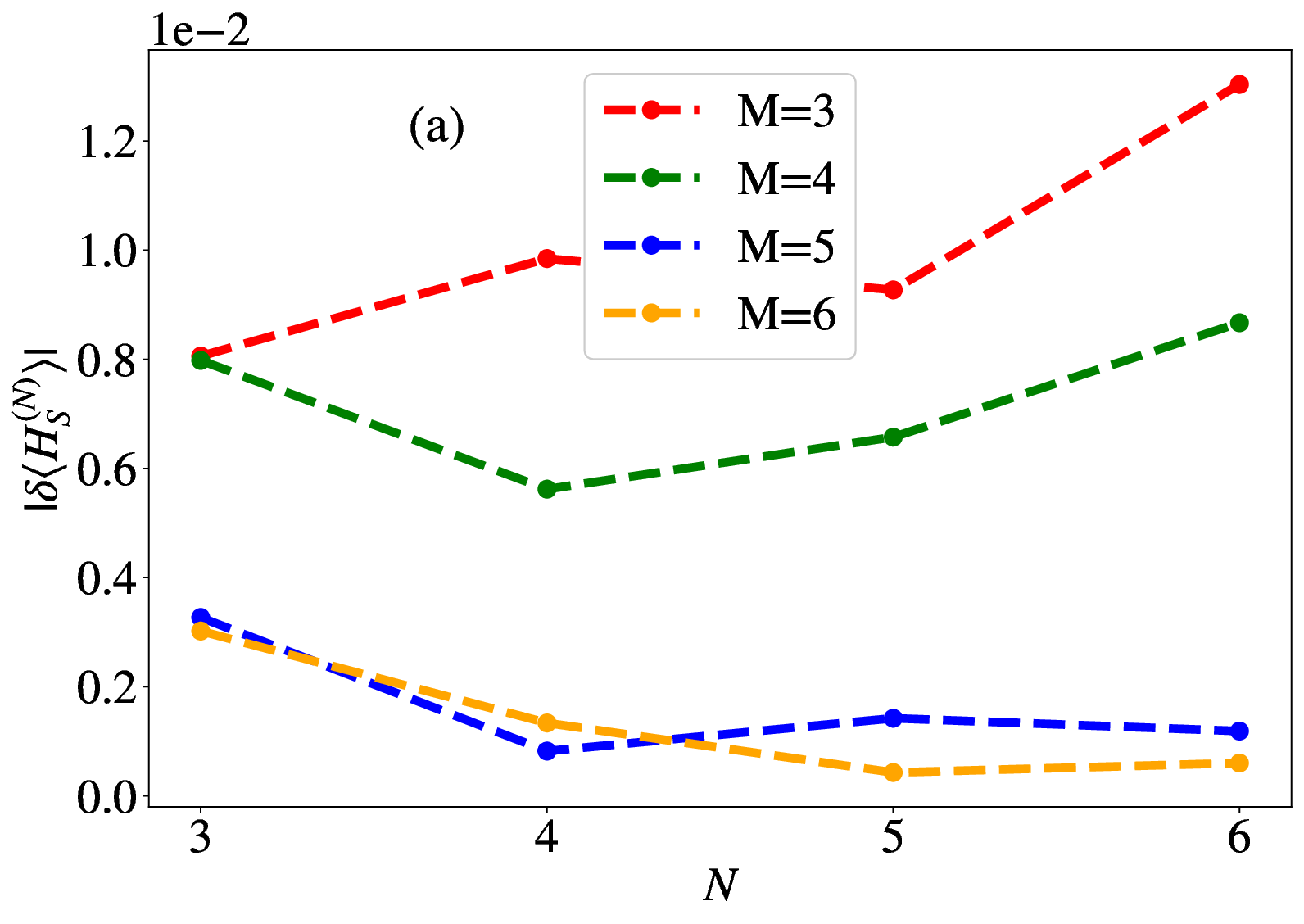} 
\includegraphics[width = \columnwidth]{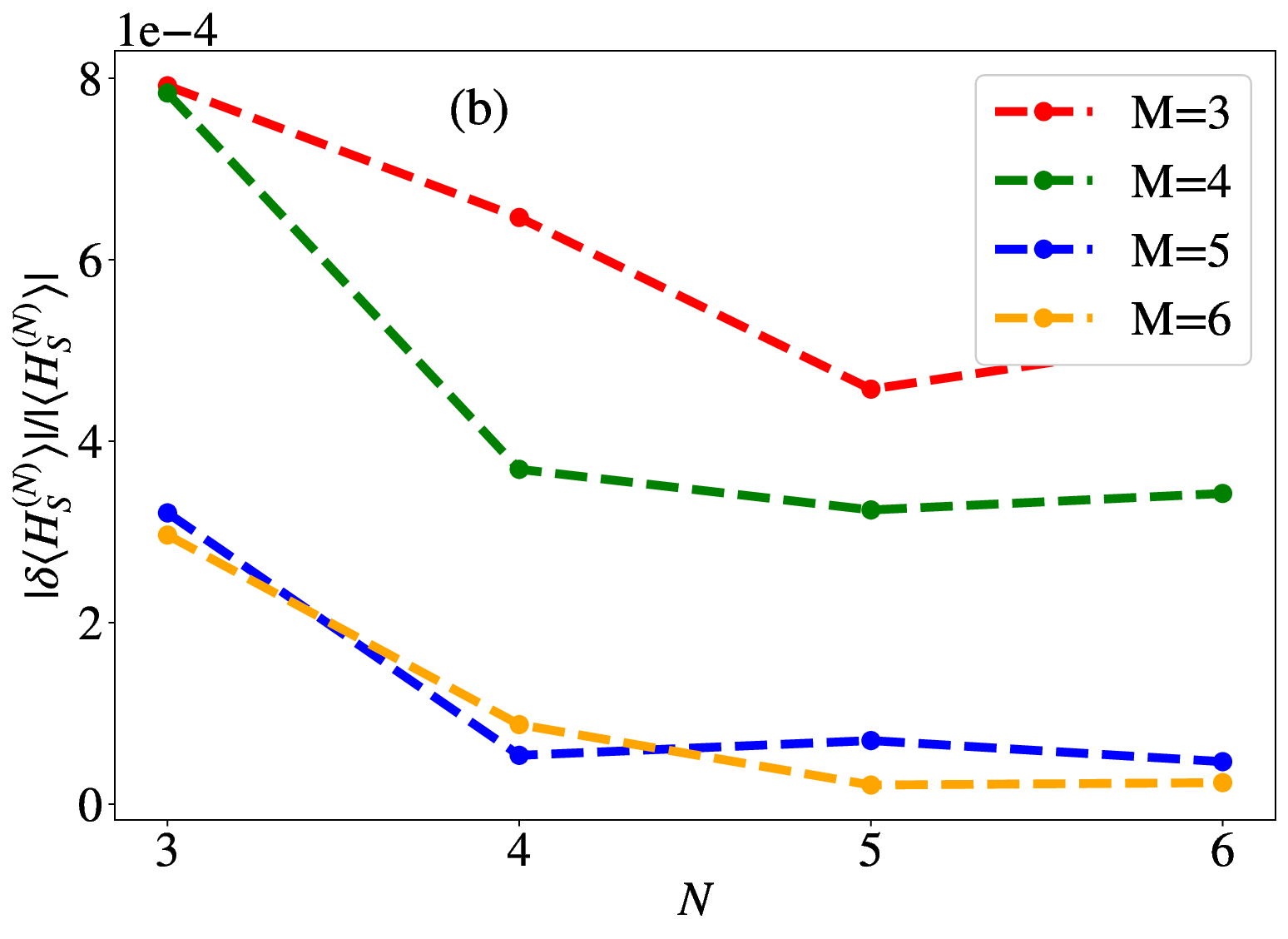} 
\includegraphics[width = \columnwidth]{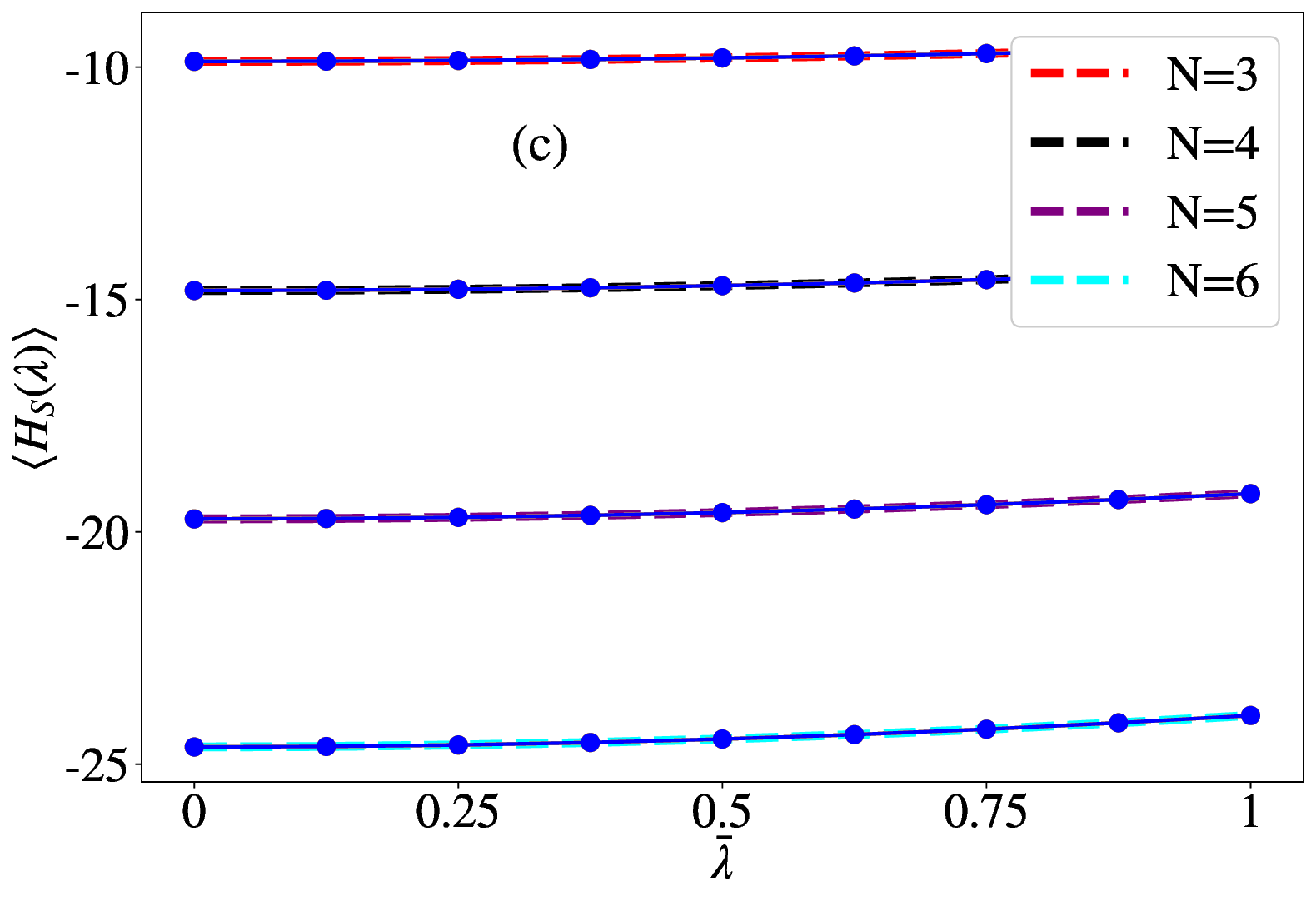} 
\includegraphics[width = \columnwidth]{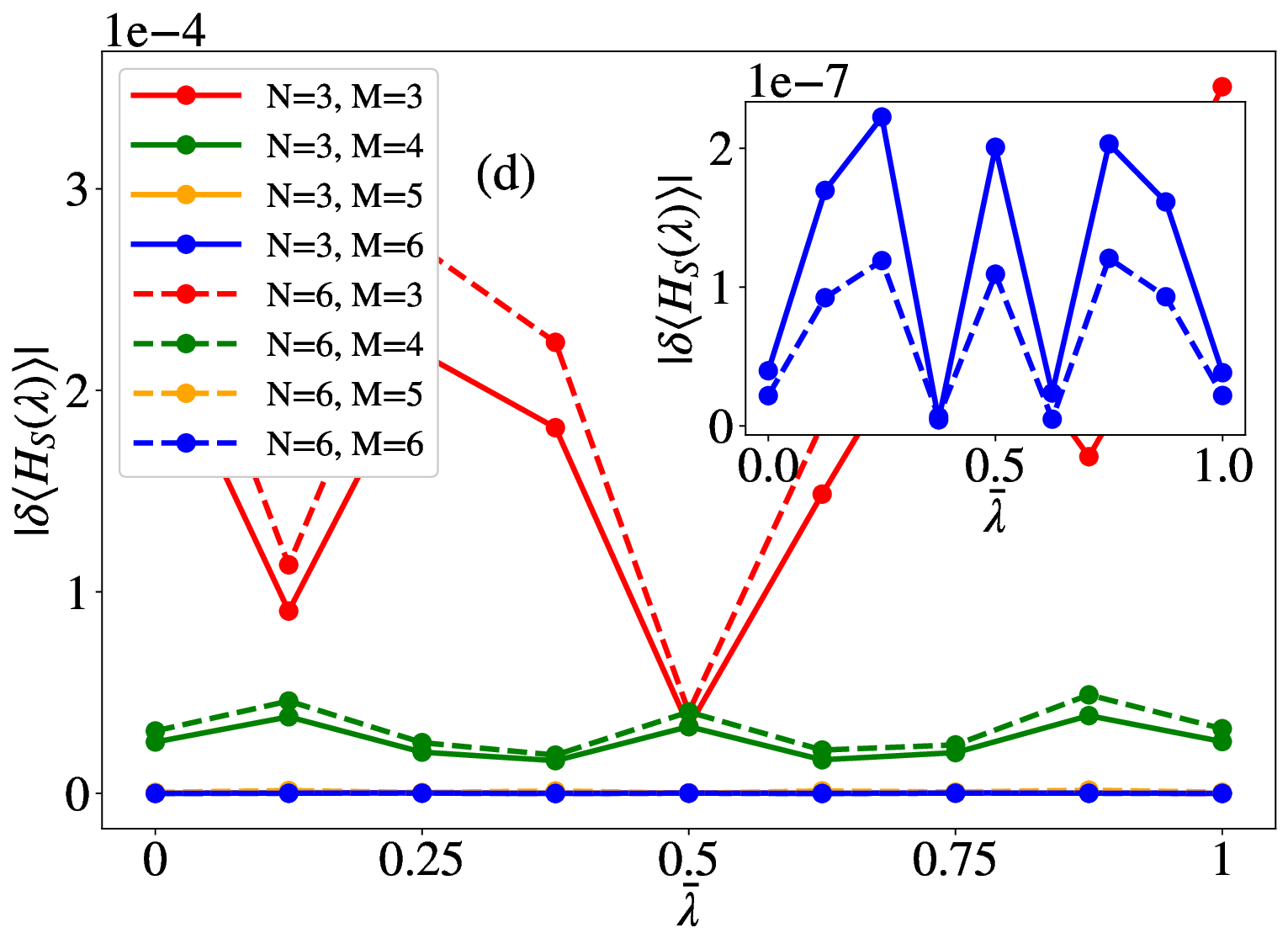} 
\caption{For the Ising model defined in the main text with $J=5g$, we show how the extrapolation process, for obtaining the full pseudomode results from an ensemble of physical models, works as a function of system size $N$. Here we are using the same parameters as in Fig.~2 in the main text: $\omega_0 = 1.2 E_{01}$, and $\lambda = 1.15g \sqrt{\Omega}$, where $\Omega = \sqrt{\omega^2 - (\gamma/2)^2}$.  We choose $\gamma = 0.37 \omega_0$ (which matches the value used in Fig.~2 of the main text when $N=5$). The top left figure (a) shows the deviation of the extrapolated total system energy from the desired value given by the full unphysical pseudomode model $|\delta \ex{H_S^{(N)}}|$, for different values of $N$ and order of fitting polynomial $M$. Similarly, the top-right figure (b) shows the same quantity scaled by the total energy.  In the unscaled quantity, (a) we see a non-monotonic trend, implying the required polynomial, and hence the error, doesn't increase indefinitely with $N$. 
 Similarly, for the rescaled extensive energy, (b), the relative error generally diminishes  with $N$.  To show this in more detail, in the bottom left (c) we show the actual fitting process as a function of $\bar{\lambda}$, $M$ and $N$, while in the bottom right (d) we show the error in the fitting function as a function. In (c) all fitting curves overlap (so the legend only indicates the original data curves, not the fits), but we do see an increase in curvature of the ground-state energy as $N$ is increased.  In (d), importantly, while for $M=3$, the size $N=3$ has a clearly lower fitting error than $N=6$, for $M=6$ they essentially have the same order error (see inset). \label{fig5}}
\end{figure*}

\section{The pseudomode method and fitting bath properties} \label{pm_appendix}

The pseudomode method was originally discussed in the context of cavity-QED by Garraway \cite{PhysRevA.55.2290}.  It was recently extended and generalized by ourselves and others \cite{lambert,Tamascelli,PhysRevResearch.2.043058, Cirio2022,Dorda,PleasanceArXiv210805755Quant-Ph2021} to deal with general bosonic and fermionic continuum environments.  The derivation of this approach, largely done in the context of modelling non-Markovian and non-perturbative environments, relies on several simple assumptions: the original free environment is Gaussian, the coupling between system and environment is linear in bath operators, and the system and environment are initially in a product state. 

The pseudomodes themselves are designed to produce the same free bath correlation functions (or, equivalently, the same free-bath power spectrum, see Fig.~1 in the main text), as some pre-defined continuum environment one is attempting to model. With the above assumptions in place, the method is only limited in accuracy by the ability of a finite number of pseudomodes to accurately reproduce some given continuum-bath correlation functions.  

Here, we are not directly concerned with their ability to mimic non-perturbative effects, as we want to minimize effects related to coherent system-environment hybridization, which may reduce the ground-state fidelity.  For the implementation here, we are also less constrained than when attempting to replicate the effects of some true environment with a given, physically relevant, spectral density. Instead, we just want to make sure that the positive-frequency part of the bath power spectrum has appropriate properties that can drive the system close to its ground state, and the negative-frequency part has a magnitude that is as close to zero as we can make it. 

Nevertheless, it is convenient to start with what might be thought of as a physical spectral density, as it makes our general analysis and control of the optimal bath properties a little easier. It also allows us to compare how well the full method works to that with just a single ancilla, similar to the approach in \cite{sciadv}. Therefore, we take the initial bath to be described by an underdamped Brownian-motion spectral density
\beq
J(\omega) =  \frac{\lambda^2 \gamma \omega}{\left[(\omega^2-\omega_0^2)^2 + \gamma^2 \omega^2\right]}.
\eeq
This has the convenient property that at zero temperature we can decompose the power spectrum into $S(\omega) = S_{a1}(\omega)+ M(\omega)$, where the first contribution is a large positive Lorentzian term,
\beq
S_{a_1}(\omega) = \frac{\lambda^2 \Gamma}{\Omega \left[(\omega -\Omega)^2 + \Gamma^2\right]},
\eeq
with $\Gamma = \gamma/2$ and $\Omega = \sqrt{\omega_0^2 - 
\Gamma^2}$. The second term, $M(\omega)$, is an infinite series of Matsubara terms, with negative amplitude.  These latter terms do directly what we need them to do: they pull the zero-frequency part of the total power spectrum to zero.  At zero temperature these latter terms become 
\beq
M(\omega) &=& \int_{-\infty}^{\infty}\;dt e^{i\omega t} M(t) \\
M(t)&=& -\frac{\gamma \lambda^2}{\pi} \int_0^{\infty} dx \frac{xe^{-xt}}{[(\Omega + i\Gamma)^2 + x^2][(\Omega - i\Gamma)^2 + x^2]}. \nonumber
\eeq

To be able to represent this with finite number of Lorentzians, and hence finite number of pseudomodes, we resort to fitting. In practice one can either fit the total $S(\omega)$ with said Lorentzians, just the $M(\omega)$ component, or fit $M(t)$ with a set of exponentials. Here we do the latter, but in more general cases there maybe some advantage to alternative fitting procedures \cite{Lambert_Bofin}.

For practical purposes here we restrict ourselves to just two fitting terms,
\beq
M_{\mathrm{fit}}(t)&\equiv&M_{a_2}(t) + M_{a_3}(t)=\lambda_2^2 e^{- \gamma_2 |t|}  +\lambda_3^2 e^{- \gamma_3 |t|} \nonumber\\
 &\simeq& |M(t)|\;,
\eeq
which in this form directly gives us the parameters used in the pseudomode master equation in  the main text which encodes the extra sign of the Matsubara contribution $M(t)$ thanks to the parameter $\bar{\lambda}\rightarrow i$. These also give rise to the two contributions plotted in the power spectrum in Fig.~1 of the main text as 
\begin{equation}
    S_{a_j}(\omega)=\bar{\lambda}^2\int_{-\infty}^\infty e^{i\omega t} M_{a_j}(t)\;dt\;,
\end{equation}
for $j=2,3$.

\begin{figure}[t!]
\includegraphics[width = \columnwidth]{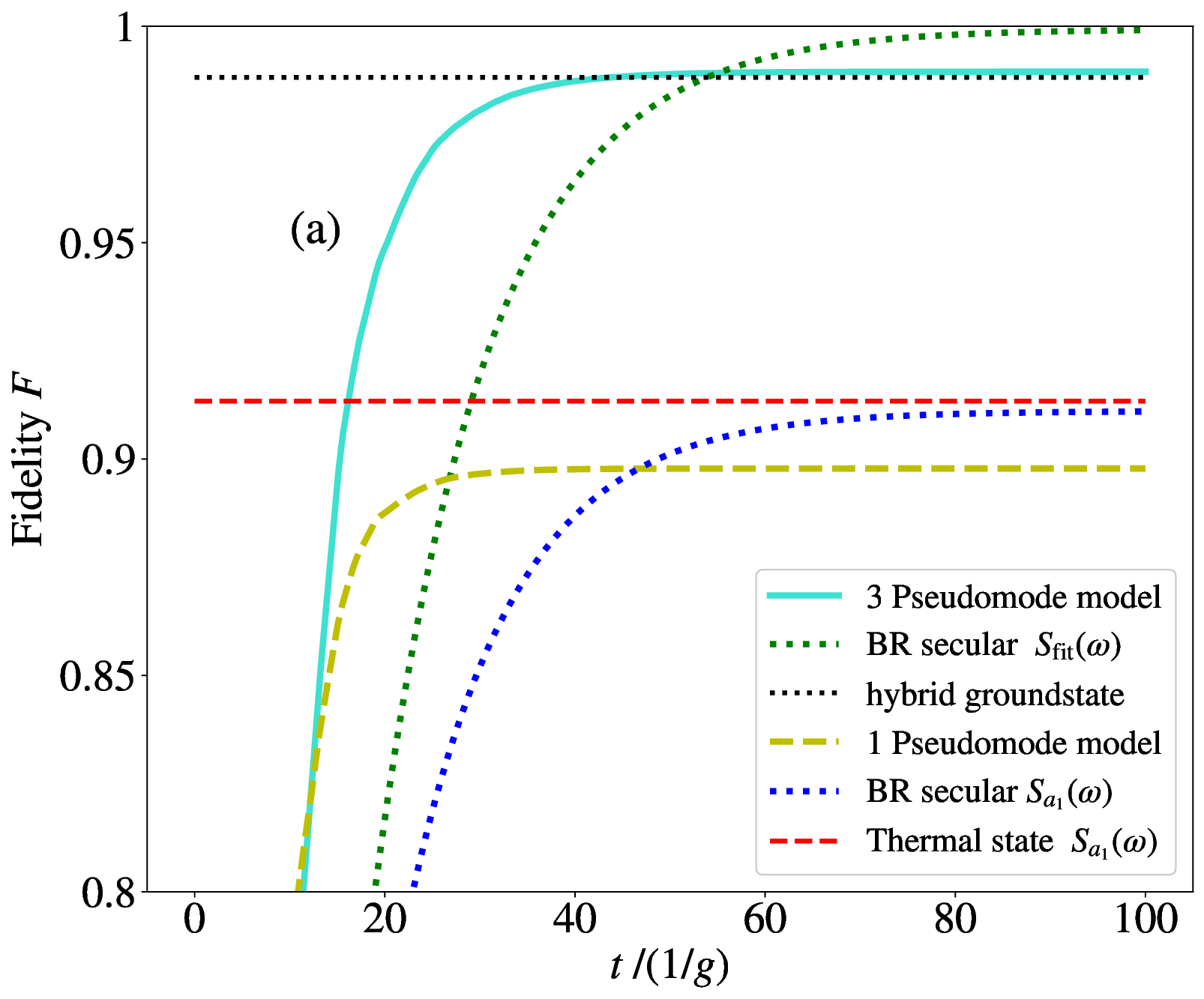} 
\includegraphics[width = \columnwidth]{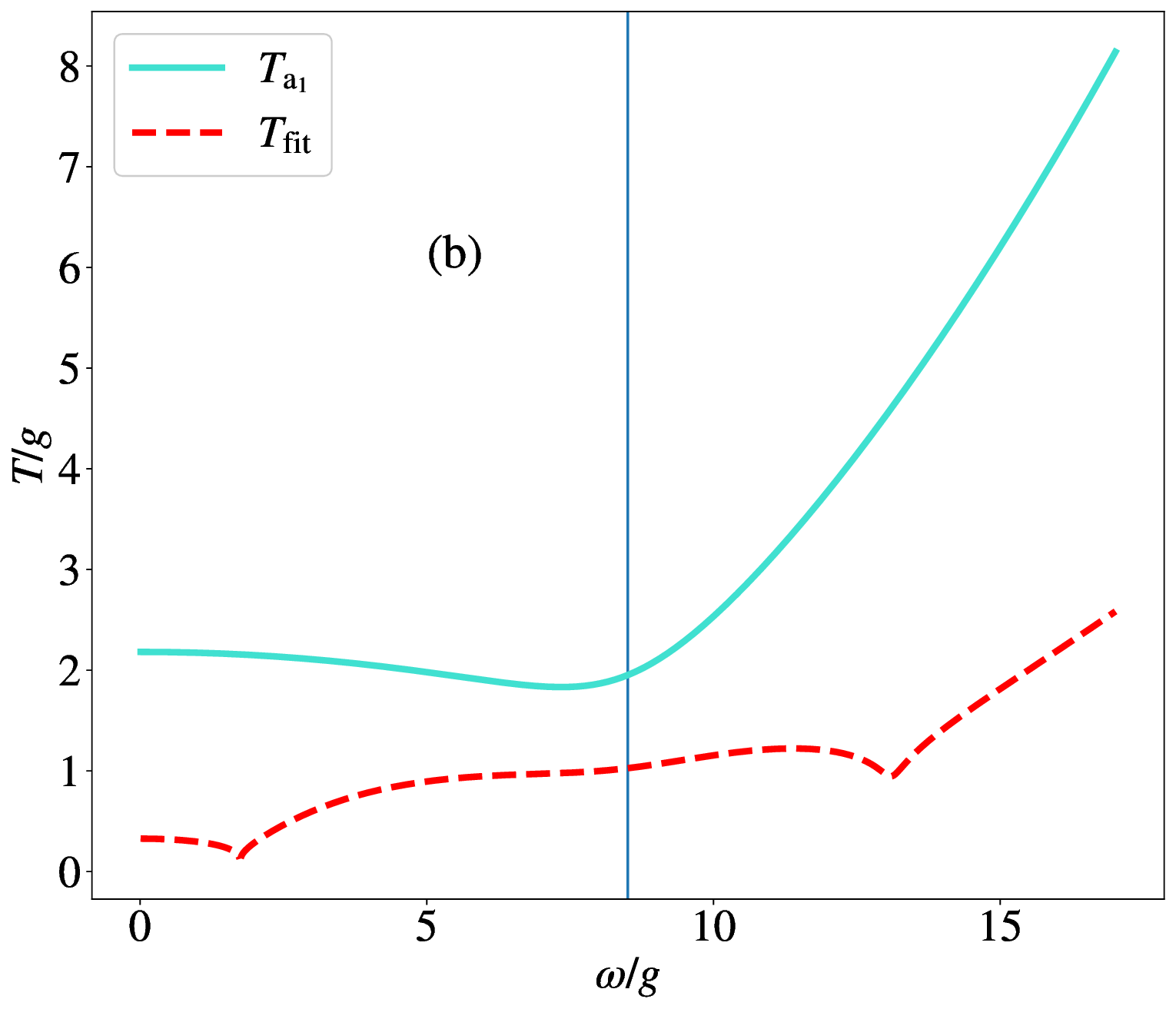} 
\caption{ (a) shows the fidelity for the same system parameters as in Fig.~2 of the main text, but with a resonant spectral density $\omega_0 = E_{01}$ to enhance the hybridization effect, lowering the overall fidelity found in the steady state.  The solid turquoise curve is the result for the full pseudomode model, while the dashed yellow curve is the result found when using just the single pseudomode $a_1$.   We see here clearly two sources of error: non-zero temperature in the single pseudomode case, and hybridization error in both the full pseudomode model and the single pseudomode case.  This can be verified by plotting the Bloch-Redfield result using $S_{\mathrm{fit}}(\omega)=S_{a_1}(\omega)+S_{a_2}(\omega)+S_{a_3}(\omega)$ (green dots) and $S_{a_1}(\omega)$ (blue dots). Furthermore, the red dashed line shows the prediction of a Gibbs thermal state for $T_{\mathrm{eff}}(\omega = E_{01})$ (but one which does not include states not connected to the groundstate by the bath coupling operator).  The black dotted line shows the fidelity of the ground state of a modified system Hamiltonian which includes just the $a_1$ mode (i.e., Eq.~1 in the main text), verifying that the dominant error in the full pseudomode is hybridization with that resonant mode. Figure (b) shows the effective frequency dependent temperature of just the single resonant mode (see main text), $T_{a_1}(\omega)$ and that of the full fit $T_{\mathrm{fit}}(\omega) = \omega \left[\log\left(S_{\mathrm{fit}}(\omega)/S_{\mathrm{fit}}(-\omega)\right)\right]^{-1}$ The blue vertical line marks the ground-state energy gap $E_{01}$.
 \label{fig6}}
\end{figure}

\bibliography{refs}

\end{document}